\documentclass[twocolumn]{aa}
\usepackage{natbib}
\usepackage{graphicx}
\usepackage{epstopdf}
\usepackage{txfonts}
\newcommand{\nodata}{ ~$\cdots$~ }    

\begin{document}
   \title{The clumpiness of molecular clouds:\\ HCO$^+$ (3--2) survey near 
Herbig-Haro objects}

\author{W.~Whyatt
          \inst{1}
          \and
          J.~M.~Girart\inst{2}
          \and
          S.~Viti\inst{1}
          \and
          R.~Estalella\inst{3}
          \and
          D.~A.~Williams\inst{1}
}

\offprints{S. Viti}

\institute{Department of Physics \& Astronomy, University College London,
Gower Street, London WC1E 6BT \\
              \email{wwhyatt@star.ucl.ac.uk}
         \and
Institut de Ci\`encies de l'Espai (CSIC-IEEC), Campus UAB,
Facultat de Ci\`encies, Torre C5 - parell 2, 08193 Bellaterra, Catalunya, Spain
	   \and
Departament d'Astronomia i Meteorologia, Universitat de Barcelona, Mart\'i 
i Franqu\`es 1, 08028 Barcelona, Catalunya, Spain
}

\date{Received  ; accepted }

\abstract
{
Some well-studied Herbig Haro objects have associated with them one 
or more cold, dense, and quiescent clumps of gas. We propose that such 
clumps near an HH object can be used as a general measure of clumpiness in 
the molecular cloud that contains that HH object. 
}
{
Our aim is to make a  survey of clumps around a sample of HH objects, and to
use the results to  make an estimate of the clumpiness in molecular clouds.
}
{
All known cold, dense, and quiescent clumps near HH objects are  anomalously
strong HCO$^+$ emitters. Our method is, therefore, to search for  strong HCO$^+$
emission as an indicator of a clump near to an HH object. The  searches were
made using JCMT (for Northern hemisphere objects) and SEST  (for Southern
hemisphere objects) in the HCO$^+$ (3--2) and also (for SEST  observations)
H$^{13}$CO$^+$ (1--0) lines, with some additional searches for methanol  and
sulphur monoxide lines. The sources selected were a sample of 22 HH objects in
which no previous HCO$^+$ emission had been detected.
}
{
We find that half of the HH objects have clumps detected  in the HCO$^+$ (3--2)
line and that all searches in H$^{13}$CO$^+$ 1-0 lines show  evidence of
clumpiness. All condensations have narrow linewidths and are
evidently unaffected dynamically by the HH jet shock.
}
{
We conclude that the molecular clouds in which these HH  objects are found
must be highly heterogeneous on scales of less than 0.1~pc. An approximate 
calculation based on these results suggests that the area filling factor of clumps 
affected by HH objects is on the order of 10\%. These clumps have gas number
densities of $\ga 3\times10^4$~cm$^{-2}$.
} 
\keywords{
ISM: abundances ---
ISM: clouds ---
ISM: molecules ---
Radio lines: ISM ---
Stars: formation 
}

\authorrunning{Whyatt, Girart, Viti, Estalella \& Williams}
\titlerunning{The clumpiness of molecular clouds}
\maketitle

%

\section{Introduction} \label{Introduction} 

The clumpy structure of molecular clouds is an important phenomenon that is
related both to interstellar dynamics and to the formation of low mass stars
occurring within these clouds. However, clumpiness is generally unresolved by
single-dish telescopes and can best be detected by interferometric observations
such as those of  \citet{Morata03,Morata05} in their studies of a portion of
the dark cloud L673 in three molecular lines. Morata et al. showed that there
were clear morphological differences for each molecular line observed.  In high
angular resolution single dish observations of TMC-1 Core D in C$_2$S lines,
\citet{Peng98} also found that clumps were abundant in the region observed, and
were resolved with sizes less than one-tenth of a parsec. Both sets of
observations implied that the clumps were transient.

\citet{Garrod06} adopted the implications of the Morata et al. (2005)
observations of clumpiness in a chemical model of a molecular cloud that was
regarded as an ensemble of transient clumps, randomly situated in both space
and time. The chemistry was followed in a time-dependent way.  These authors
showed that this cloud model is able to account in a natural way (based on the
time-dependence of the chemistry) for the observed characteristic structures
and morphological differences between different species at both low and high
angular resolution.

In the present paper, we present an alternative observational  approach to the
study of clumpiness within molecular clouds. This approach relies on the
observational result that some Herbig-Haro objects are found to have associated
with them small dense condensations of gas first observed in enhanced emission
in lines of HCO$^+$ and NH$_3$, e.g.: HH 1/2 
\citep{Torrelles92,Girart02,Girart05};  HH 7-11  \citep{Rudolph88,Dent93}; HH 34
\citep{Rudolph92};   NGC 2264G \citep{Girart00}.  In spite of the nearby HH
shocks, the regions of enhanced HCO$^+$ emission are found to be quiescent;
their temperatures are close to 10 K, and their linewidths are narrow (on the
order of 1~km~s$^{-1}$). These regions are much denser than the typical cloud
density, and the observations suggest H$_2$ number densities on the order of
$10^5$~cm$^{-3}$.

The condensations associated with HH objects have also been studied in the case
of HH 2 in lines of 12 molecular species  \citep{Girart02},  in addition to
HCO$^+$ and NH$_3$. These observations confirm that the condensations are as
dense and cold as indicated by the previous HCO$^+$ and NH$_3$ observations. 
This work also made clear that the chemistry of these condensations is distinct
from that of the ambient cloud.  \citet{Viti06}   showed that this chemical
anomaly was shared by at least five other HH objects. While subsequent and more
detailed studies of HH 2  \citep{Girart05,Lefloch05} showed that the region
ahead of the shock is more complex than had been previously considered and that
some dynamical interaction with the HH shock may be occurring, nevertheless,
significant parts of the condensations near HH 2 are confirmed to be quiescent,
cold and dense.

Are the condensations associated with HH objects similar to the clumps directly
observed by interferometric means in L673? If so, then we may regard HH objects
as a probe of clumpiness in molecular clouds, or at least those regions surrounding the HH objects. It is therefore important to
determine the extent of clumpiness as revealed by HH objects. Since
observations made by many authors over the last three decades all confirm that
the condensations associated with HH objects are strong emitters in HCO$^+$, we
use strong HCO$^+$ (3--2) emission as a signature of a dense clump in the
vicinity of an HH object. We report here the results of an HCO$^+$ (3--2)
survey of condensations associated with 22 HH objects. The aim of this work was
to determine a measure of the clumpiness in molecular clouds containing HH
objects.

In $\S~2$ we describe the HCO$^+$ observations made in the HH objects, and
in $\S~3$ we give the results of the data obtained. In
$\S~4$ we give a brief discussion of clumpiness deduced from these
observations of HCO$^+$ emitting condensations near HH objects, and make some
conclusions about the nature of clumpiness in molecular clouds that  contain HH
objects.


\section{Observations}

The northern hemisphere observations were carried out with the JCMT telescope
on Mauna Kea between December 2002 and August 2003.  The JCMT observations were
done in typical band 4 weather ($\tau_{\rm 230GHz}\sim 0.12$--0.2), with the
heterodyne A3 receiver centred at 267.56~GHz for the HCO$^+$ (3--2) line.
Additional observations of the CH$_3$OH ($5_n-4_n$) rotational line were taken
in the position of the strongest HCO$^+$ emission for HH~29. The full width at
half maximum (FWHM) beam size is $\simeq20''$ for these two lines. As the cold,
quiescent regions produce narrow linewidths $\leq 1$~km~s$^{-1}$, the minimum
bandwidth of 125~MHz was used to produce a velocity resolution of $\sim
0.12$~km~s$^{-1}$. This bandwith allows for the detection of the two strongest
lines of CH$_3$OH ($5_{-1}$--4$_{-1}$, $5_{0}$--$4_{0}$). The typical
integration time was 4~min per offset for the HCO$^+$ observations and 27 mins
per offset for the methanol observations in order to obtain a baseline rms of
0.15~K and 0.05~K respectively. Main beam efficiency for the HCO$^+$ (3--2)
line was 0.69.

The southern hemisphere observations were carried out with the SEST on La Silla
in January 2003. In the SEST observations we used dual receiver capability to
observe simultaneously the HCO$^+$ (3--2) and the H$^{13}$CO$^+$ (1--0) lines.
We used an integration time of 6 min per offset, which yielded an average
baseline RMS noise of 0.2 and 0.05~K, at a velocity resolution of 0.2 kms$^{-1}$, for
the HCO$^+$ (3--2) and H$^{13}$CO$^+$ (1--0) lines, respectively. The FWHM beam
size is $\simeq20''$ and $54''$ for these two lines, respectively. In certain
cases, but only if observations of HCO$^+$ (3--2) were successful, additional
simultaneous observations of CH$_{3}$OH ($2_n$--$1_n$) and SO
(6$_{5}$--$5_{4}$) were undertaken.  To reach an average baseline RMS noise of
0.03~K, at a velocity resolution of 0.1~km~s$^{-1}$, required a typical
integration times of 18 mins per offset. Main beam efficiency for the  HCO$^+$
(3--2),H$^{13}$CO$^+$ (1--0), CH$_{3}$OH ($2_n$--$1_n$) and SO
(6$_{5}$--$5_{4}$) lines were 0.42, 0.75, 0.72 and 0.52, respectively.

A successful detection has been defined as a 3$\sigma$ detection at a velocity resolution of 0.2~km/s, a tentative detection as a signal that is above 2$\sigma$. The data were processed using `CLASS' from the `GILDAS' software package.  Gaussian fits were attempted within an appropriate velocity window close to the V$_{LSR}$ of the region. These velocities are typically within 1~km/s of the known V$_{LSR}$ for the surrounding cloud from the literature. 

\begin{table*}
\caption{Source List. JCMT and SEST. General Information. Column (7) lists the distance between the exciting source and the HH Object. Column (8) lists the distance between the proposed HH Object and the peak emission of HCO$^+$ (3--2) seen. `N/A' indicates that the exciting source is not known and no values can be deduced, whereas a `--' indicates that no HCO$^+$ emission was seen.}
\label{General}
\centering
\begin{tabular}{l cc l l c  c c l }
\hline\hline
 		& RA (J2000) 	& Dec (J2000) 			& $\!\!$V$_{LSR}$ 		&        		& Distance 	& Dist. source 	 & Dist. HH &Proposed  \\
 Object   	&  (h m s) 	& ($^{\rm o}$ $'$ $''$) 	& $\!\!\!\!$(km~s$^{-1}$)	&  Location 	& (pc) 		& to HH (pc) 		& 
to HCO$^+$ (pc) 
 & Exciting Source$^a$ \\
\hline
HH~267 &03 24 03.0 	&$+$31 00 29 	& 4.5 	& L1448	  	& 300 	&2.53 	&--			&L1448C\\
HH~268 &03 24 22.1 	&$+$30 48 11 	& 4.5 	& L1448	  	& 300 	&1.35 	&--			&L1448~IRS1\\
HH~337A&03 28 26.3 	&$+$31 25 53 	& 4.0 	& NGC~1333  	& 300 	&0.99	&0.03		&333 star\\
HH~278 &03 26 59.4 	&$+$30 25 58 	& 7.0 	& L1455/L1448 & 300 	&2.47 	&--			&L1448NB  \\
HH~366 &03 48 29.8	&$+$32 55 36 	& 5.0 	& Barnard 5 	& 300 	&1.28 	&0.01 		&B5 IRS 1 \\
HH~427 &03 30 37.7 	&$+$30 21 56 	& 6.0 	& Barnard 1 	& 300  	&N/A 		&N/A		&Not known \\
HH~211 &03 43 56.8 	&$+$32 00 50 	& 9.2 	& IC~348    	& 300 	&0.08 	&0.01 		&HH 211-mm \\
HH~462 &03 54 05.0 	&$+$38 10 35	&$-3.5$	& Perseus 	& 300 	&0.01	&0.04		&IRAS~03507+3801 \\
HH~362A&04 04 24.1 	&$+$20 20 41 	& 7.2 	& L1489	   	& 140 	&0.20	&--			&IRAS~04106+2610 \\
HH~464 &04 10 42.4 	&$+$38 07 39	&$-3.5$	& L1473	   	& 350 	&0.07	&0.05			&PP~13N \\
HH~276 &04 22 07.3 	&$+$26 57 26 	& 6.5 	&Taurus-Auriga& 140	&1.80	&--			&IRAS~04189+2650  \\
HH~29  &04 31 27.6 	&$+$18 06 24 	& 6.6 	& L1551	  	& 140 	&0.11	&0.01		&L1551~IRS 5       \\
HH~240 &05 19 40.7 	&$-$05 51 44 	& 8.5 	& L1634	  	& 460 	&0.28 	&0.11		&IRAS 05173$-$0555 \\
HH~43  &05 38 10.4 	&$-$07 09 25 	& 8.5 	& L1641	  	& 460 	&0.57 	&0.17		&IRAS~05355$-$0709C\\
HH~38  &05 38 21.8 	&$-$07 11 38 	& 8.5 	& L1641	  	& 460 	&1.06  	&0.07		&IRAS~05355$-$0709C\\
HH~272 &06 12 48.4	 	&$-$06 11 21 	&11.0	& L1646	  	& 830 	&0.46 	&0.08		&IRAS~06103$-$0612 \\
HH~47C &08 25 33.0 	&$-$51 01 37 	& 6.0 	& Gum Nebula	& 450 	&0.26 		&--			& HH~46/47~IRS      \\
HH~75  &09 11 38.5 	&$-$45 42 28	&$-0.9$	& Gum Nebula	& 450 	&-- 		&--			&IRAS~09094$-$4522 \\
HH~49  &11 06 00.1 	&$-$77 33 36 	& 5.3 	& Cha I	  	& 160 	&2.69 		&--	 		&Cha-MMS1         \\
HH~52-53&12 55 06.4	&$-$76 57 46 	& 4.0 	& Cha II    	& 165 	&1.39	&--			&IRAS~12496$-$7650 \\
HH~54  &12 55 49.5 	&$-$76 56 08 	& 4.0 	& Cha II    	& 165 	&1.90	&--			&IRAS~12496$-$7650\\
HH~77  &15 00 49.0 	&$-$63 07 46	&$-5.0$	& Circinus  	& 700 	&0.63 	&0.06		&IRAS~14564$-$6254 \\

\hline
\end{tabular}
\begin{list}{}{}
     \item[$^{a}$] 
References: 
HH29: \citet{Moriarty06}; HH38--43: \citet{Stanke00}; 
HH49: \citet{Reipurth96}; HH52--53 \& HH54: \citet{Knee92}; 
HH47: \citet{Sahu89}; HH75 \& HH77: \citet{Cohen90}; 
HH211: \citet{McCaughrean94}; HH240: \citet{Davis97};
HH267: \citet{Bally97};  HH268: \citet{Eisloffel00}; HH272: \citet{Carballo92};
HH276: \citet{Wu02}; HH278: \citet{Eisloffel00}; HH337A: \citet{Bally96b};
HH362: \citet{Alten97}; HH366: \citet{Bally96a}; HH427: \citet{Yan98}; 
HH462 \& HH464: \citet{Aspin00}
\end{list}
\end{table*}

The sample was selected according to the following criteria: (i) in order
to avoid emission from dynamically affected material, we selected HH
objects located at the end of the jet, (ii) usually several HH objects are
found in the same region (e.g. there are 36 known HH objects in NGC~1333);
we selected one HH object per outflow (except for HH~38 and HH~43), to
avoid the bias of one region being particularly clumpy with respect to the
norm, (iii) we limited our sample to HH objects that are clearly embedded
in a dark molecular cloud and, to our knowledge, outside obvious star
forming regions, (iv) where possible, we chose HH objects for which CCD
images were available from the literature, in order to aid the
interpretation of the spatial distribution of the emission.  The offset
positions from the HH objects were chosen on the basis of known HCO$^+$
clumps ahead of well studied HH objects such as HH~2, which are usually
found at a distances of $\la0.1$~pc downstream of the HH object. However it is important to note that clump does not have to be directly ahead of the HH object. In any case the direction and orientation of the jet is often not known with great certainty, nevertheless the clump is ahead of the HH object in the sense that it has yet to be reached by the jet, and thus has not been dynamically effected by the HH object.

The sources were observed in the HCO$^+$ (3--2) line for numerous offsets
around a sample of HH objects. The specific offset positions were kept flexible during our observations so if emission was seen we would be able to adapt the observing procedure to allow us explore this extent of this emission, within the constraints of our available time for each source.

Table 1 shows the list of the HH objects, their position, the $v_{\rm LSR}$ of the molecular cloud, the name of the region and its distance, the distance between 
proposed exciting source and the HH object, the distance between the HH object 
and the peak emission seen in HCO$^+$ (3--2) and finally the suspected exciting 
source of the HH object.

\section{Results}

In total, we observed 12 northern hemisphere objects with the JCMT and a
further 10 southern hemisphere sources with the SEST (see Table~\ref{General}
for details). In this section we will briefly describe our source sample and
report our detections, if any, for each source, before analysing the emission
seen. Figures 1 to 12 show the HCO$^+$ (3--2) spectra (and H$^{13}$CO$^+$ for
the SEST sample) of the sources with detected emission. 
Figure~\ref{fig_metanol} shows detections of the CH$_3$OH (2$_n$--1$_n$) line.
No detections were made in the SO line.

\subsection{Details of sources}

\begin{figure} 
\centering
\includegraphics[width=7.5cm]{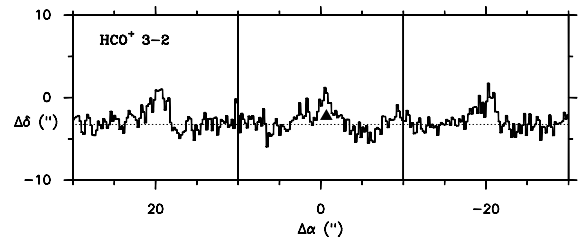}
\caption{Spectra of the HCO$^+$ (3--2) emission HH~29, observed with
JCMT. The horizontal and vertical axes shows the offset position (with
respect to the HH object) of the observed spectra. 
The temperature scale for the spectra is the same for all the figures showing 
the HCO$^+$ (3--2), ranging from $T_{\rm mb}=-1.0$ to 2.0~K. 
The HCO$^+$ spectra have a velocity resolution of 0.18~km~s$^{-1}$. 
These spectra have the same velocity range, with a width of 16~k~ms$^{-1}$ centered at the systemic velocity of the molecular cloud (this also
applies for the H$^{13}$CO$^+$ spectra shown in other figures). 
The triangle marks the position of the HH object.
}
\label{29}
\end{figure}

{\it HH~29:} A bright bow shock situated in the large outflow 0.11~pc from the
binary, L1551~IRS~5 first seen by \citet{Herbig74}. More recent observations
have identified a second outflow from neighbouring   L1551~NE with the same
orientation as the IRS~5 flow and which is easily confused with the original
flow. Because of this, it is unknown which source is powering HH~29, although
it is now more commonly thought to be L1551~NE due to deep narrow band images
and proper motion measurements of HH~29 \citep{Moriarty06}. The HCO$^+$ (3--2)
line is detected in all the three offsets (Figure~\ref{29}). This yields to a
lower limits for the emission scale of about 0.04~pc.

{\it HH~38 \& HH~43:} Discovered by \citet{Haro53}, both are part of the same
HH flow driven by IRAS 05355$-$0709C, also known as  HH~43 MMS1, which also
powers HH~64 \citep{Stanke00}. HH~43 is closer to the source with a projected
distance of 0.57~pc, whilst the terminating object, HH~38, is a further 0.5~pc
downstream.  We detect HCO$^+$ (3-2) in both HH~38 and HH~43, in a single
position, suggesting a compact emitting region. However, the detection towards
HH 43 is $47''$ south (0.1~pc) of knots E and F.

\begin{figure} 
\centering
\includegraphics[width=7.5cm]{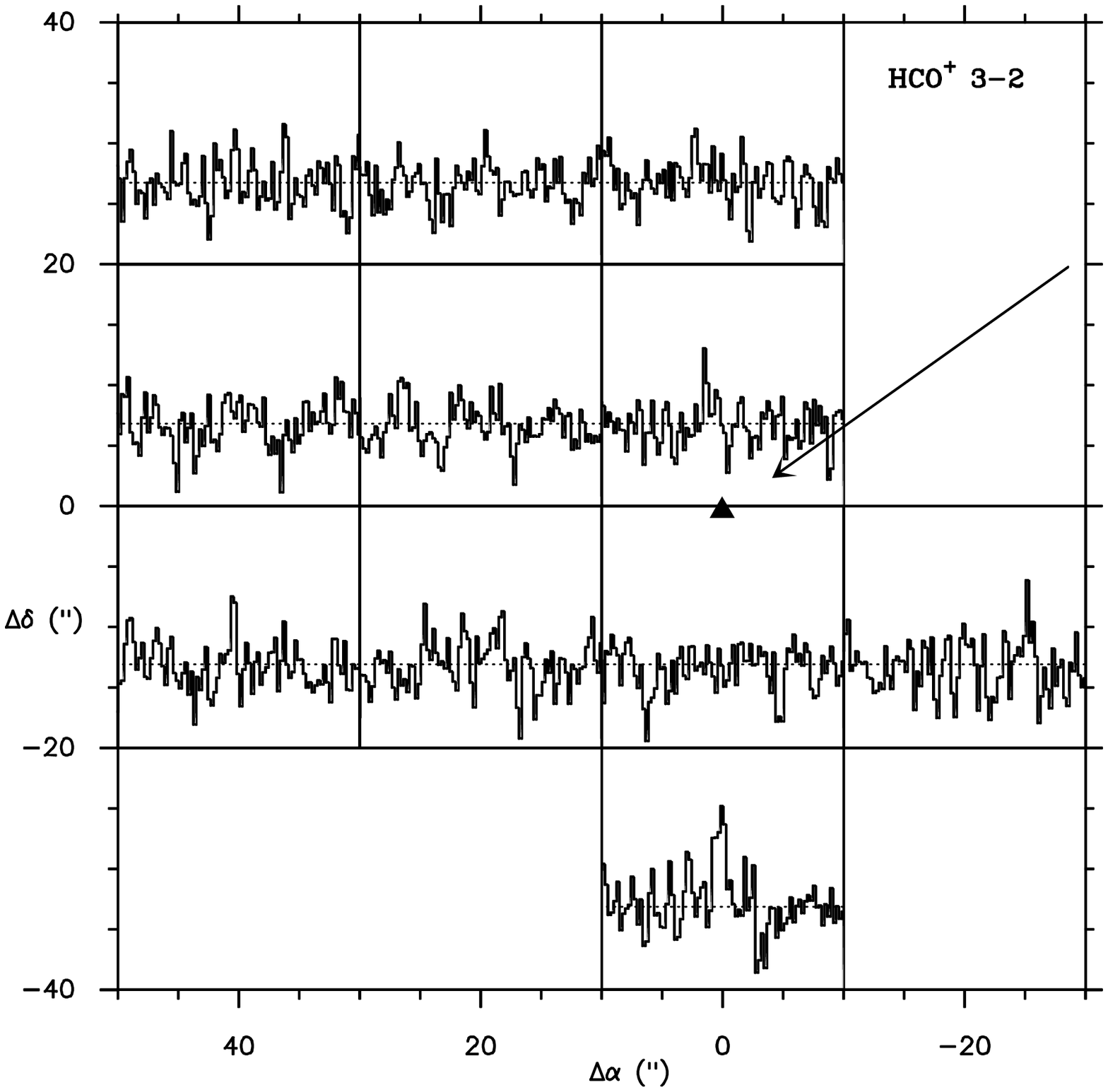}
\includegraphics[width=7.5cm]{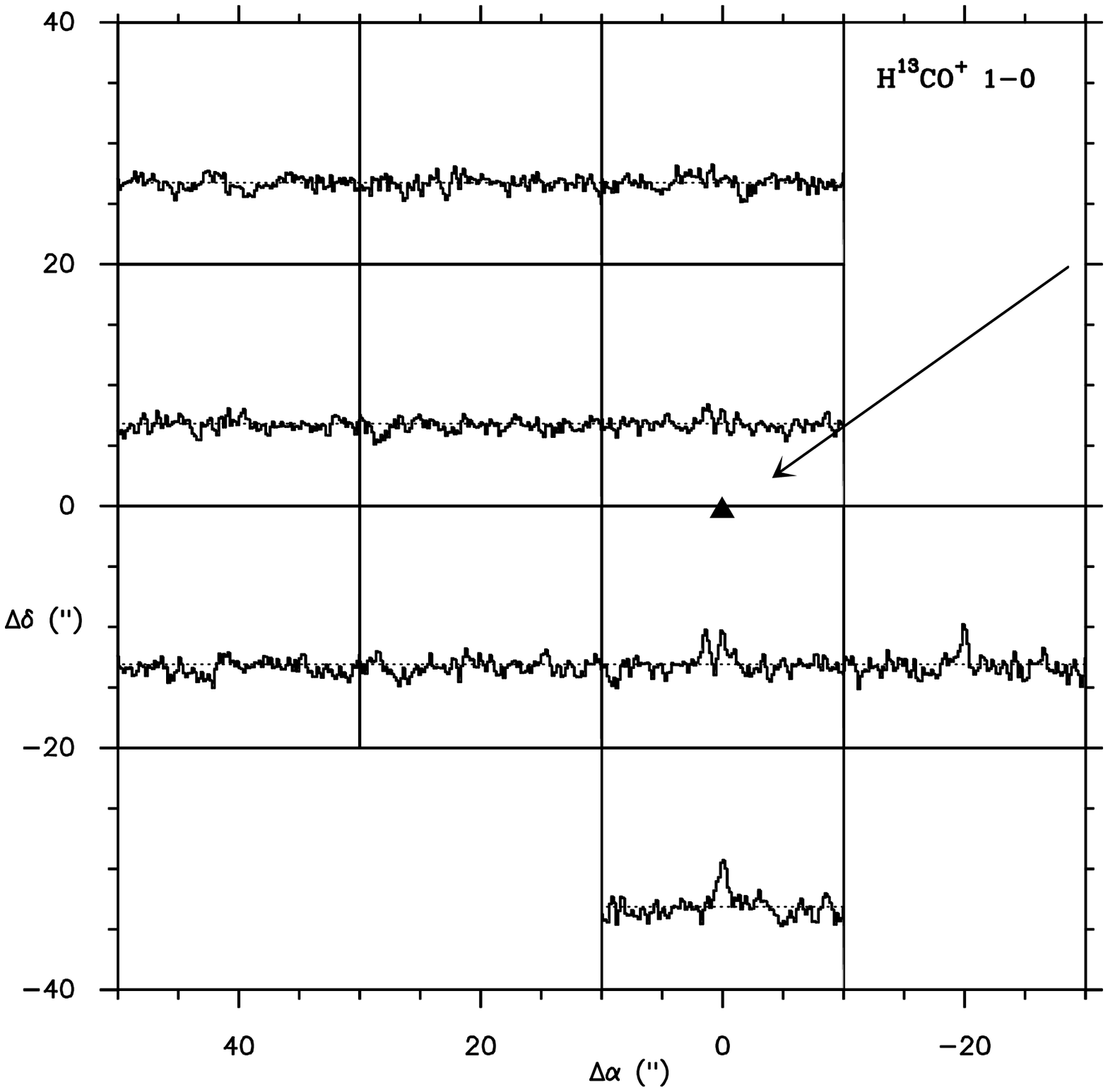}
\caption{Spectra of the HCO$^+$ (3--2) (top) and H$^{13}$CO$^+$ (1--0) emission
(bottom) ahead of HH~38, observed with SEST. 
The temperature scale is the same for all the figures showing the  
H$^{13}$CO$^+$ spectra, ranging from $T_{\rm mb}=-0.5$ to 1.0~K. 
The H$^{13}$CO$^+$ spectra have a velocity resolution of 0.15~km~s$^{-1}$. 
The arrow shows the direction of the outflow and the triangle marks the position 
of the HH object.
} 
\label{38}
\end{figure}

\begin{figure*}
\includegraphics[width=7.5cm]{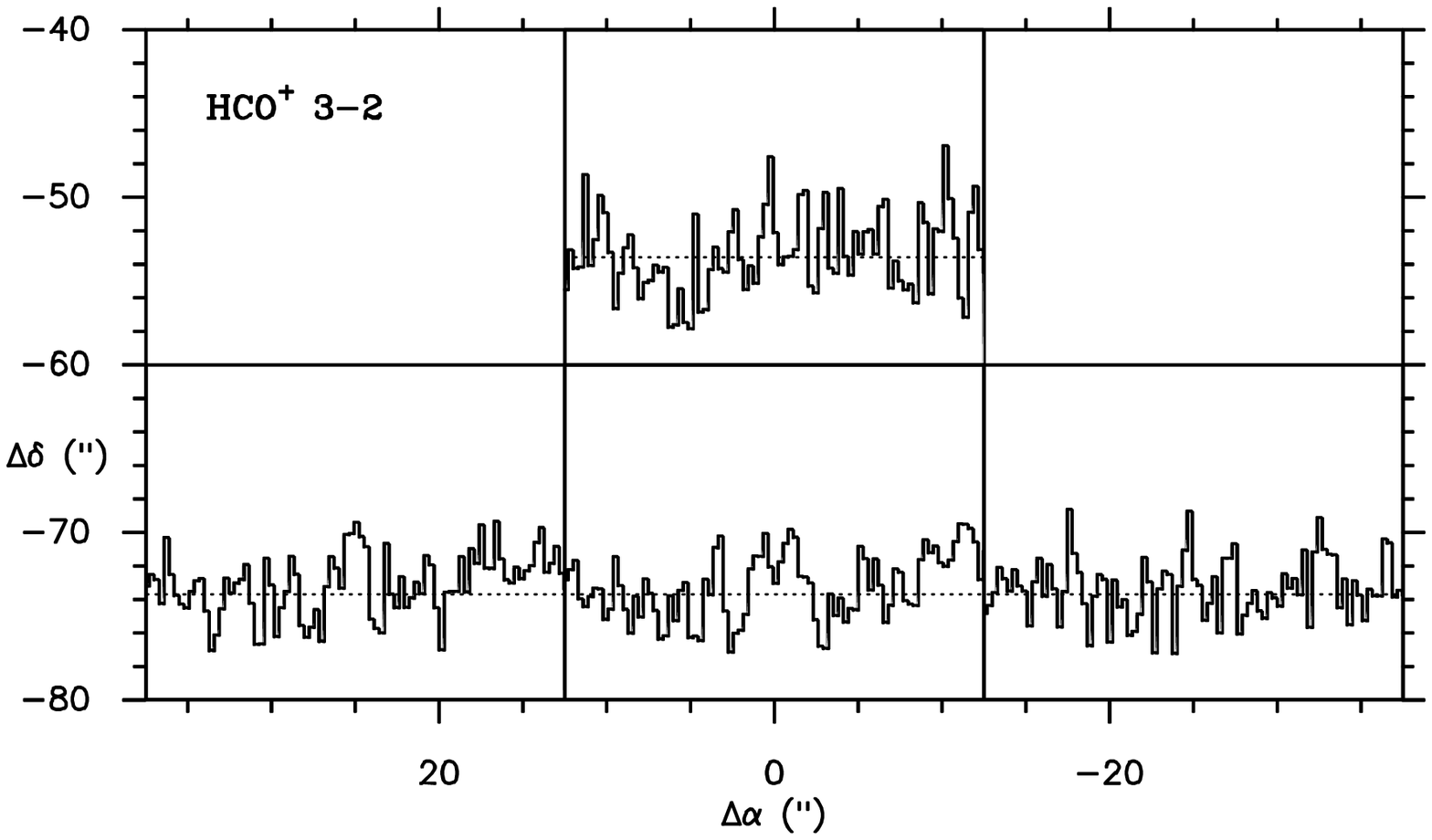}
\hspace*{0.3cm}
\includegraphics[width=7.5cm]{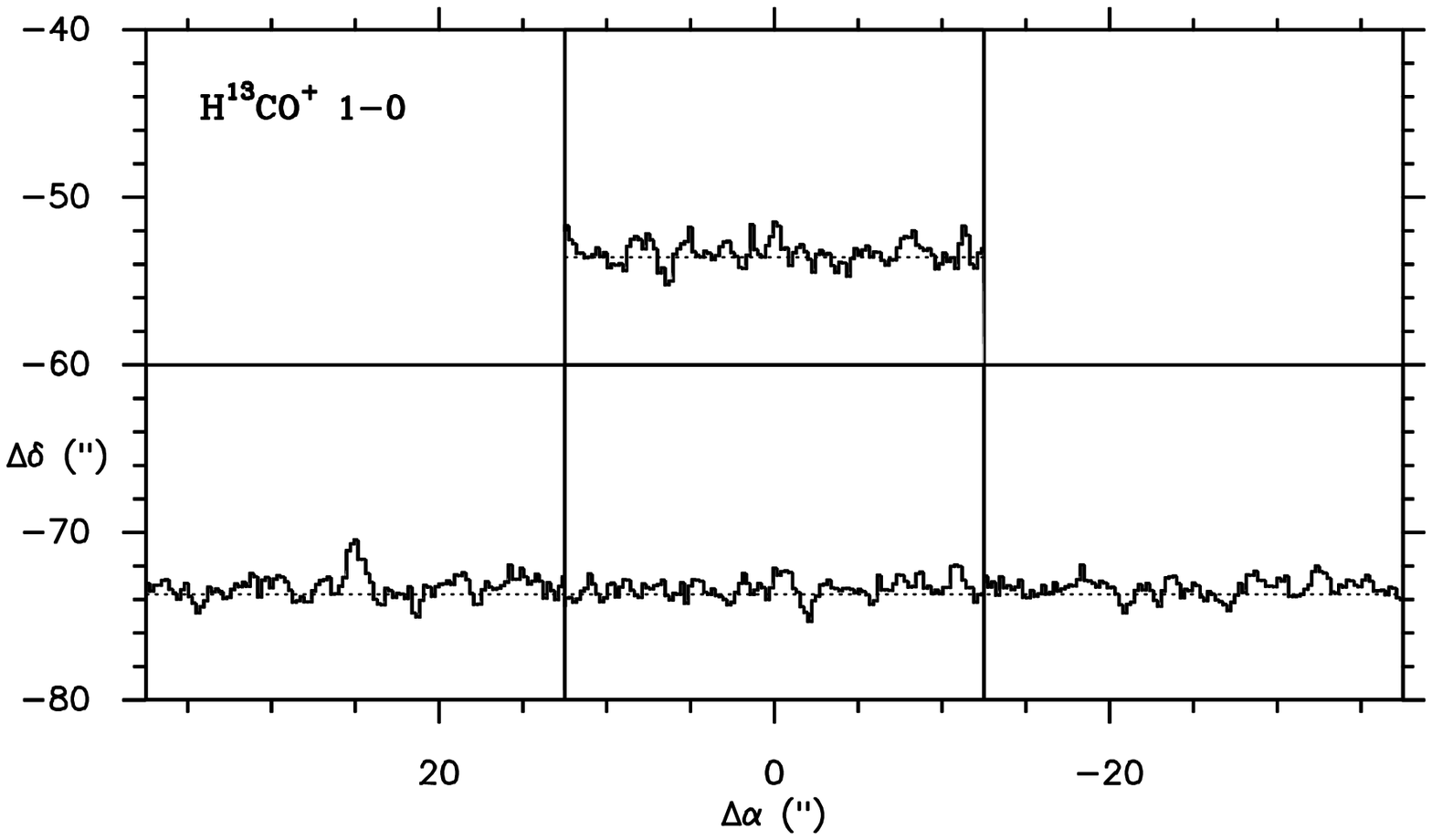}
\caption{SEST spectra of the HCO$^+$ (3--2) and H$^{13}$CO$^+$ (1--0) ahead of 
HH43. The position offsets are with respect to the brightest knot in HH 43.
}
\label{43}
\end{figure*}

{\it HH~47C:} Along with HH~46, HH~47 is a large, well collimated bipolar
outflow embedded in the Bok globule ESO~210-6A. Unlike the brighter HH~47A and
the older and larger HH~47D, HH~47C is powered by a faint counter-jet and is
located approximately  0.26~pc from the central source. It was first discovered
by \citet{Schwartz77b}. We report no detections in the eleven positions
observed.

{\it HH~49:} Discovered by \citet{Schwartz77a} and along with its close
companion HH~50, they are the brightest shocks in the Cha I cloud. Located
south of the reflection nebula Ced~110, they trace a large bow flowing out from
that region. The observations of \citet{Reipurth96} and  \citet{Bally06}
indicate that Cha-MMS1 is the driving source - located 0.47~pc upstream from
HH~49. We report no detection in any of the 22 observed positions.

{\it HH~52--53 \& HH~54}: Again first identified by \citet{Schwartz77a}, these
sources seem to form a chain heading in a north east direction as part of a bow
shock driven by IRAS~12496$-$7650 (0.67~pc southwest).  However, \citet{Knee92}
proposed that HH~52--53 and HH~54 are independent HH objects and that they are
powered by closer exciting source. HH 54 is spatially coincident with a weak IR
source, IRAS~12522$-$7640, which has also associated a monopolar, blueshifted
outflow. HH~52--53 is possibly powered by another weak IR source,
IRAS~12522$-$7641. We found no HCO$^+$ around these HH objects.

{\it HH~75:} This is a complex chain of knots located at the southern edge of
the irregular cloud S114.  Observations of associated nebulous stars by
\citet{Herbst75} indicate the closer Gum Nebula is the host to the objects - as
such  \citet{Cohen90}, put forward IRAS 09094$-$4522 as the source located
1.1~pc away. No emission was detected.

{\it HH~77:} First identified by \citet{Reipurth88}, this is a small curved
object near several possible powering IRAS sources. \citet{Cohen90} put forward
IRAS 14564$-$6254 (located 0.4~pc to the WNW) as the powering source due a
streamer connecting the two objects. Interestingly the orientation of the
bow-shock puts the source in the opposite direction to the IRAS source. This
implies that the flow may be colliding with dense clumps and is in the process
of flowing around it \citep{Schwartz78}.
There are 4  detections ($\ge$ 3~$\sigma$) in the HCO$^+$ 3--2
line that are clearly detected in the H$^{13}$CO$^+$ 1--0

\begin{figure} 
\centering
\includegraphics[width=7.5cm]{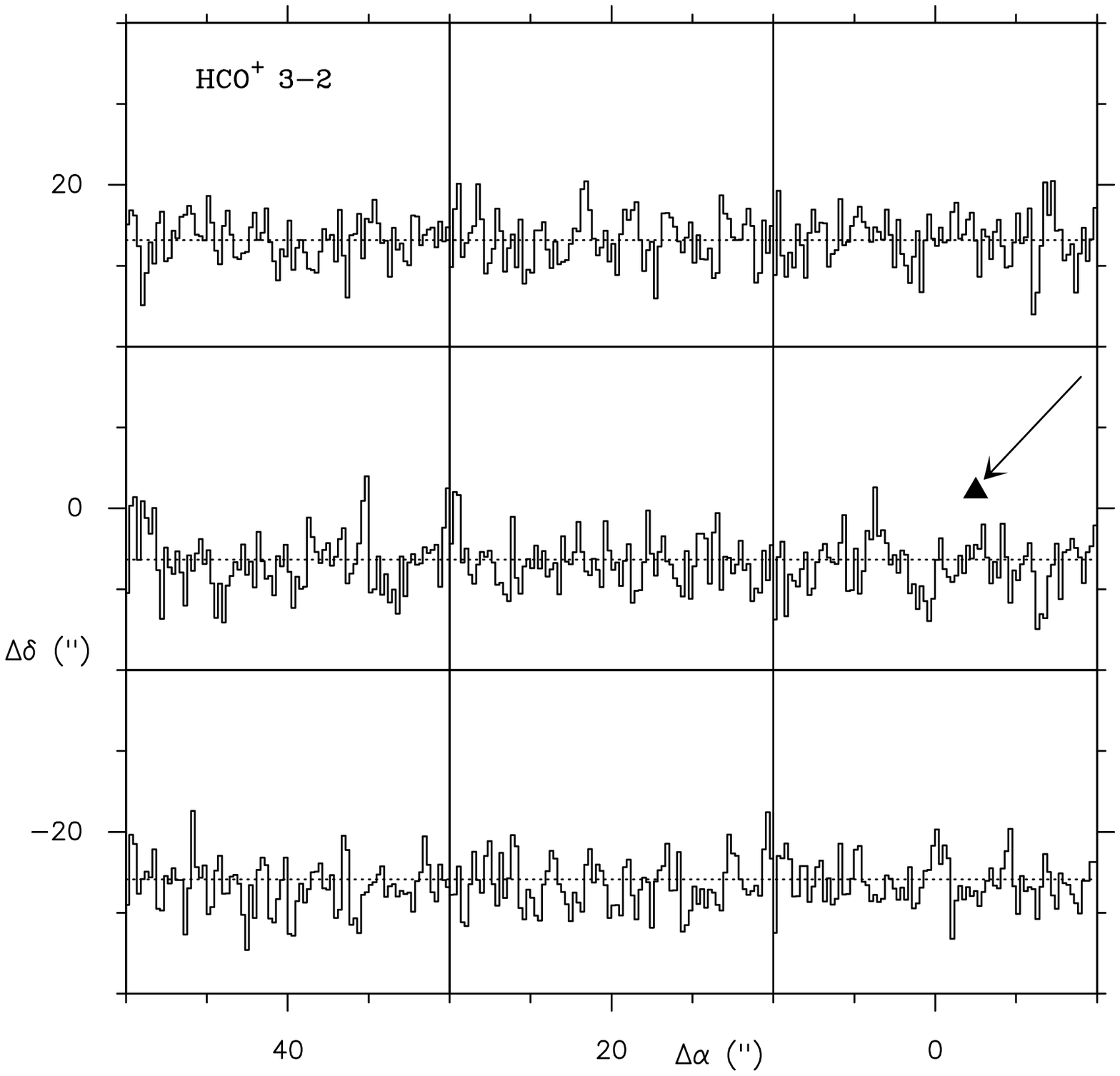}
\includegraphics[width=7.5cm]{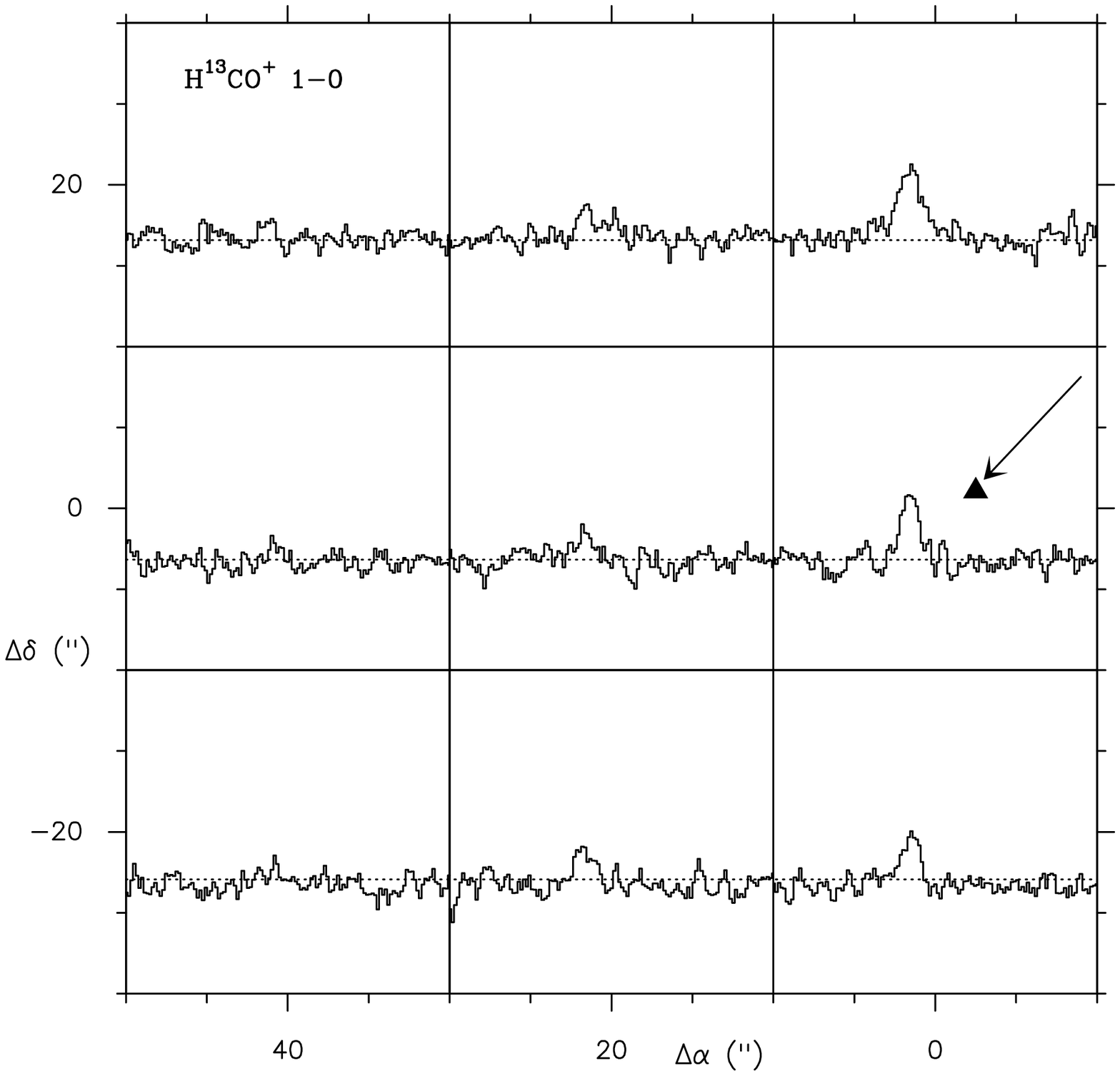}
\caption{Spectra of the HCO$^+$ (3--2) (top) and  H$^{13}$CO$^+$ (1--0)
(bottom) emission east of HH~77, observed with SEST.}
\label{77}
\end{figure}

{\it HH~211:} Discovered by \citet{McCaughrean94}, HH~211  features a
relatively small but highly collimated, bipolar jet (0.16~pc long), seen in
molecular hydrogen (IR), and a molecular outflow, both powered by a cold sub-mm
source. For HH~211, HCO$^+$ (3--2) emission was seen
in each position observed (although falling in intensity with distance from
the bow shock) - as such, emission will extend outside the region observed, so
we calculate a minimum size of 0.1~pc. 

\begin{figure} 
\centering
\includegraphics[width=5.5cm]{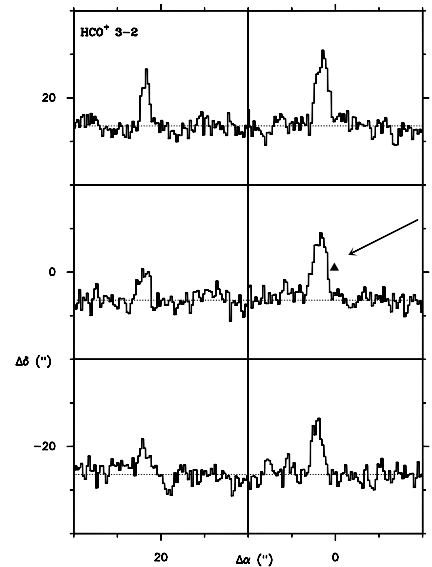}
\caption{Spectra of the HCO$^+$ (3--2) emission west of HH 211, observed with
the JCMT.}
\label{211}
\end{figure}

{\it HH~240:} First seen by \citet{Cohen80} under the name RNO 40, it is a
chain of HH knots with a bright central core. \citet{Hodapp95} and
\citet{Davis97} infrared H$_{2}$ observations indicate that HH 240 forms a
highly symmetric bipolar outflow with the much fainter HH~241, placing
IRAS~05173$-$0555 as the source 0.28~pc away.   HH~240 appears to have a
relatively small clump in HCO$^+$ (3--2) of approximately 0.07~pc in length. 

\begin{figure} 
\centering
\includegraphics[width=9.5cm]{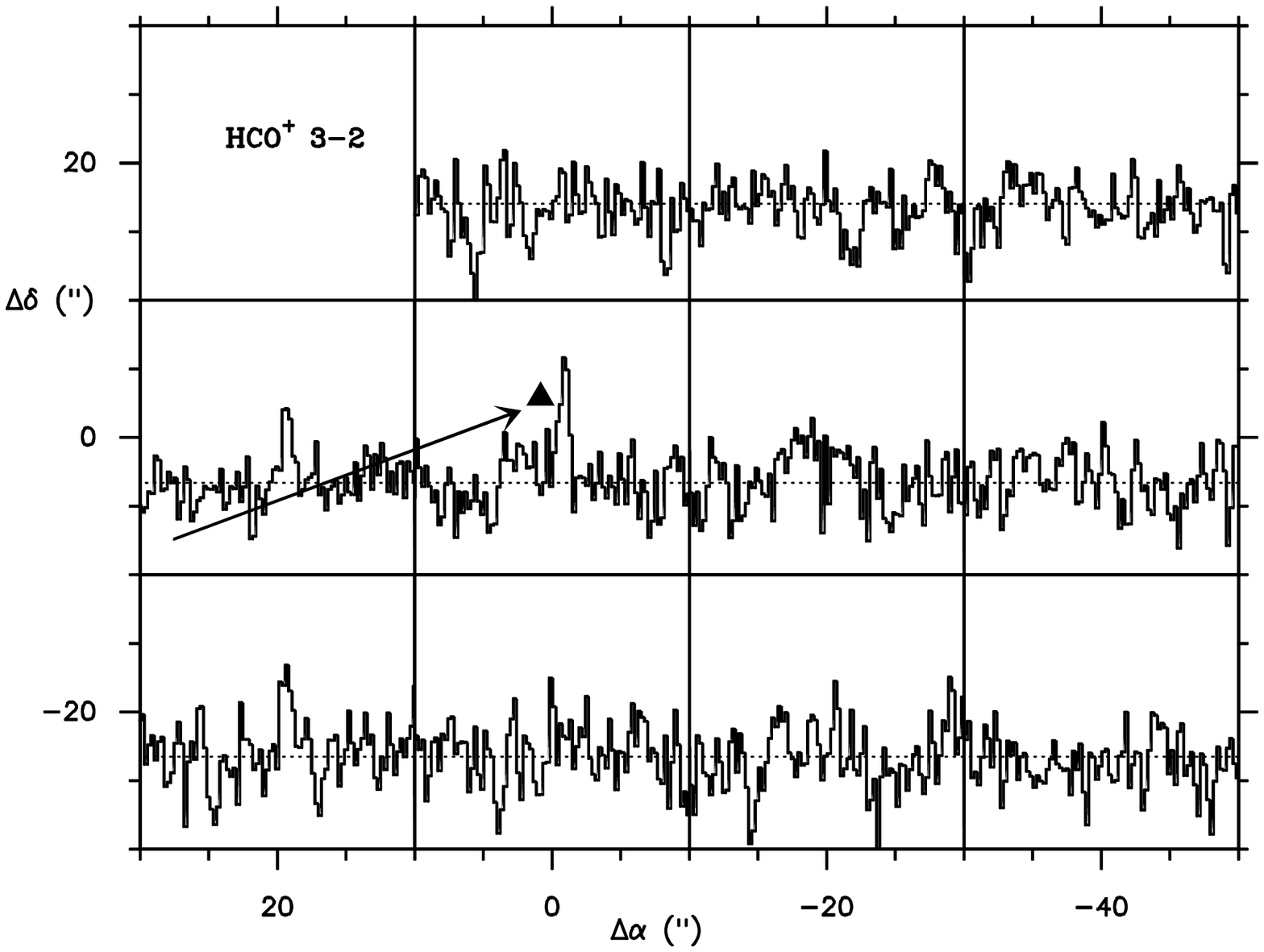}
\includegraphics[width=9.5cm]{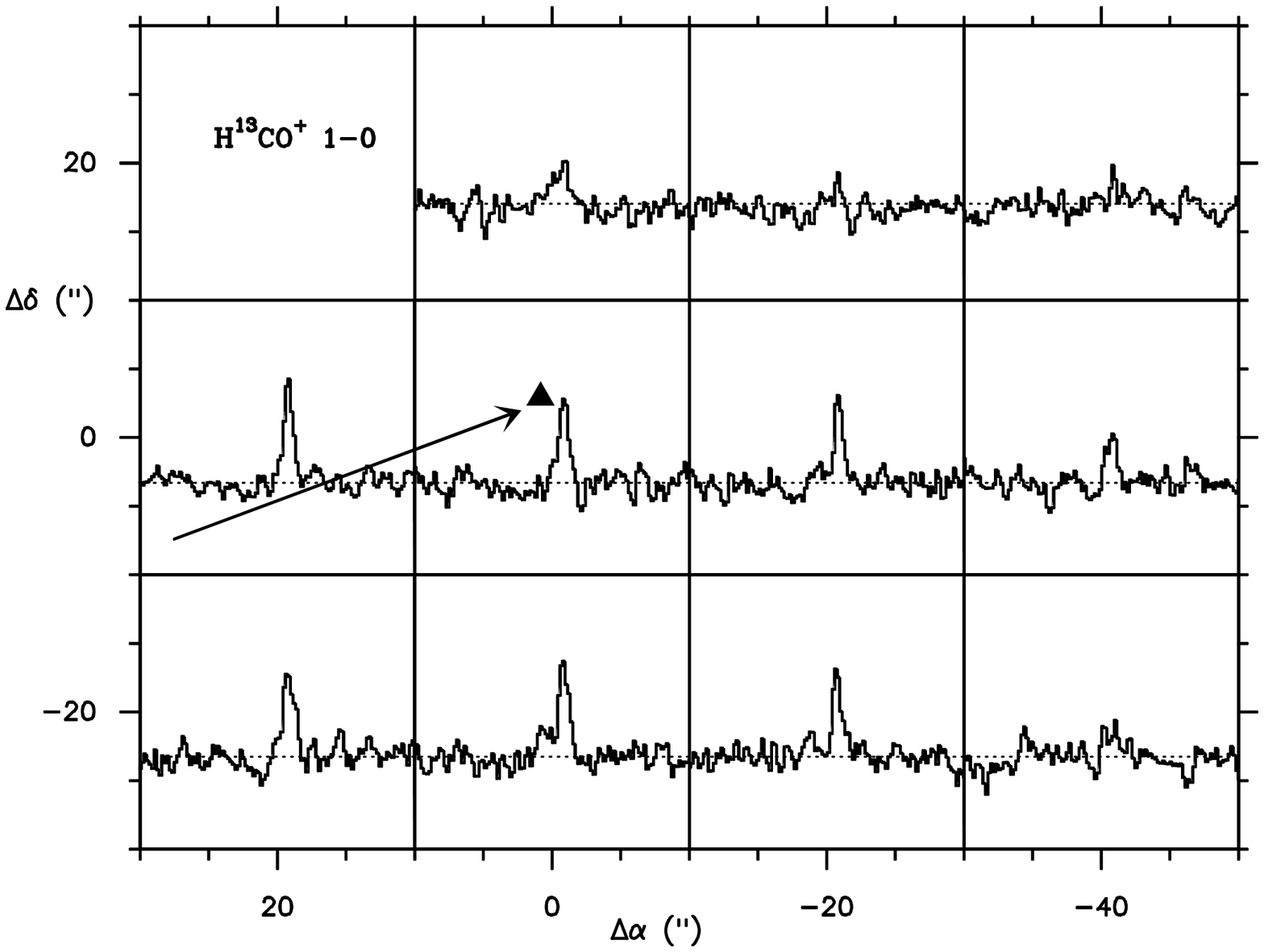}
\caption{Spectra of the HCO$^+$ (3--2) (top) and H$^{13}$CO$^+$ (1--0) emission
(bottom) ahead of HH~240, observed with SEST. }
\label{240}
\end{figure}

{\it HH~267:} First discovered by \citet{Bally97}, HH~267 is an extended
cluster of knots. L1448C is the likely source of HH~267, as it lies on the axis
of its outflow and because the terminal velocity of this outflow is consistent
with the velocity of HH~267. The outflow from L1448C is believed to be 0.71~pc
long, whilst if it powers HH~267 this extends the flow to a total length of
2.5~pc. We report no detections for this source.

{\it HH~268:} Again discovered by \citet{Bally97}, HH~268 is a series of faint
knots located 1.35~pc from the powering source L1448~IRS~1 \citep{Eisloffel00}.
We report no detections  for this source. 

{\it HH~272:} \citet{Carballo92} discovered this chain of 11 knots located near
the reflection nebula GGD~17 in the Mon R2 Region.  HH~272 is 0.72~pc to the
north from its powering source IRAS 06103$-$0612 (also known as Bretz 4). We
observed a 0.24~pc by 0.24~pc region surrounding knots I and H. The 
H$^{13}$CO$^+$ line is detected in several positions, but the HCO$^+$ line is 
detected in three positions. The emission seen to the SW is 0.18 pc from the emission seen close to the HH object and so may be tracing another unrelated clump or extended emission.

\begin{figure} 
\centering
\includegraphics[width=7.5cm]{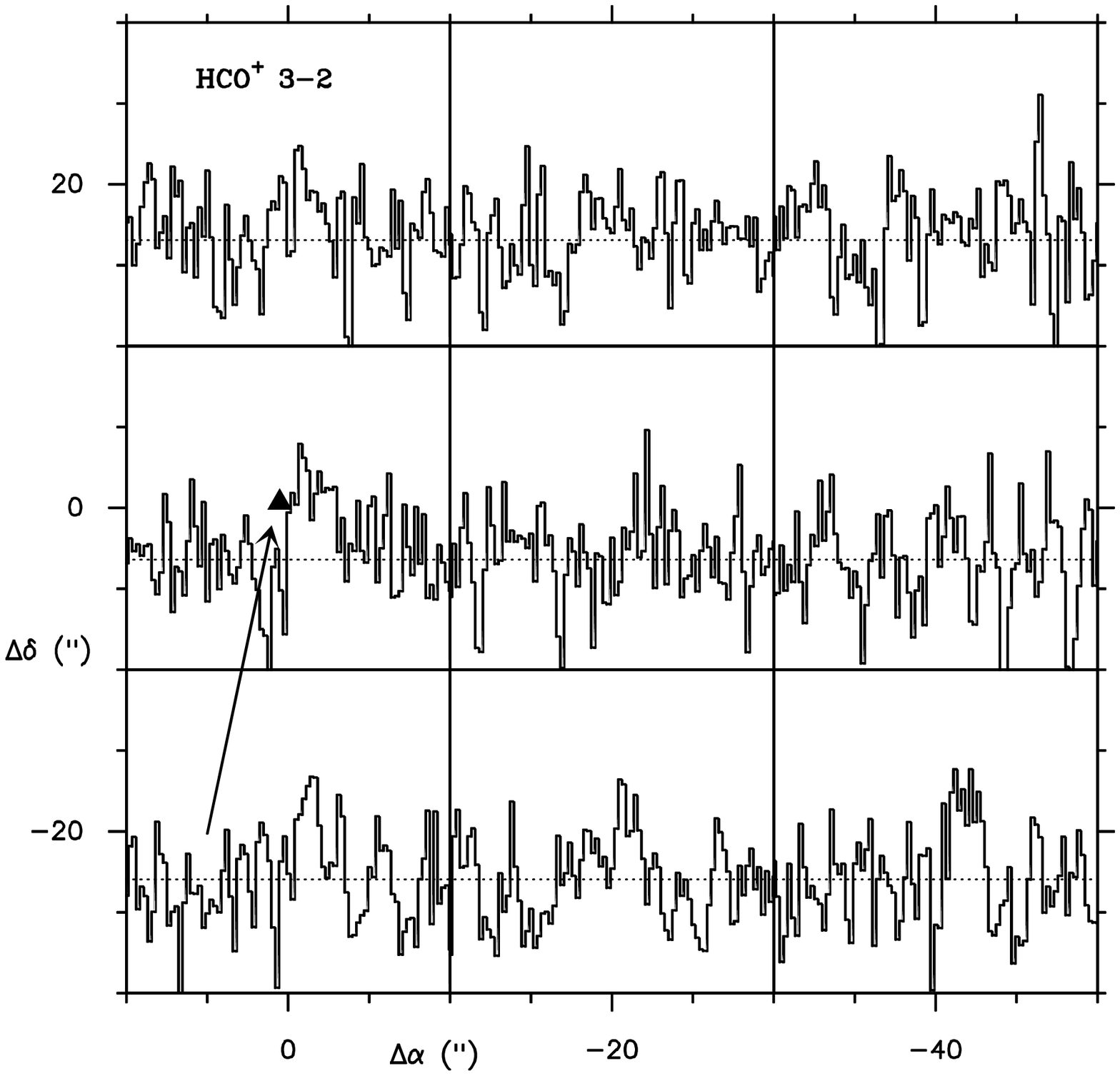}
\includegraphics[width=7.5cm]{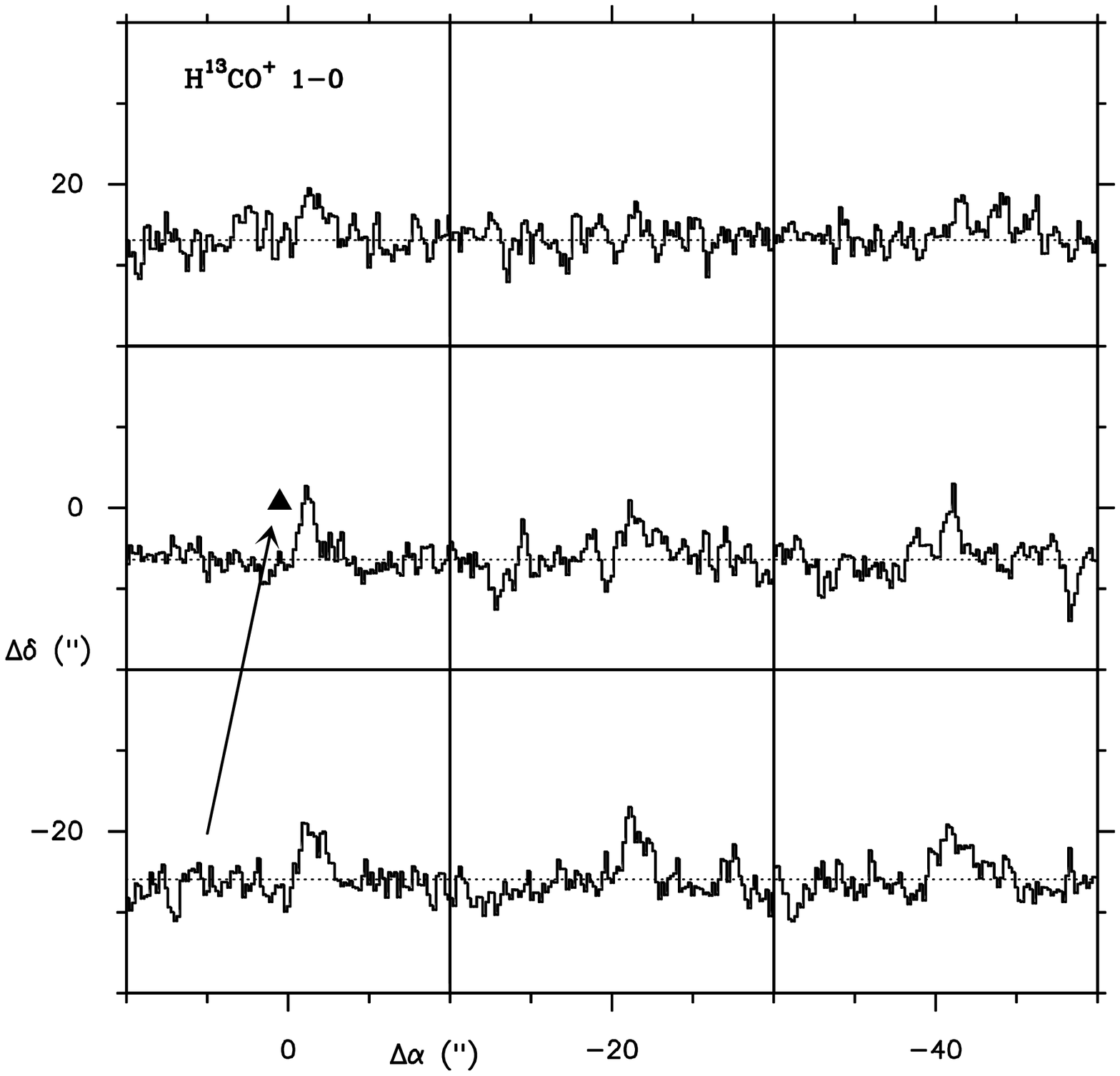}
\caption{Spectra of the HCO$^+$ (3--2) (top) and H$^{13}$CO$^+$ (1--0) emission
(bottom) ahead of HH~272, observed with SEST.}
\label{272}
\end{figure}

{\it HH~276:} Originally discovered by \citet{Eisloffel98}, it is a series of 4
knots (A to D) that are perpendicular to the far larger Tau B outflow.  It is
centered roughly 60'' (0.04~pc) northeast of FS Tau B, with no known exciting
source. The knots broaden to the SE implying the source is NW of HH~276.
\citet{Wu02} put forward IRAS 04189+2650 as the powering source, 1.8~pc away.
We obtained no detections for this source.

{\it HH~278:} this is a large diffuse object, consisting of at least 3 knots.
First seen by \cite{Bally97},  it is located roughly equidistant from L1448 and
L1455. The exciting source has not been confirmed - although the outflow from
L1448-N(B) \citep{Eisloffel00} is the most likely source as the HH~278 and the
outflow are aligned - despite the large 2.47~pc distance between them. No
detections were made.

\begin{figure}
\centering
\includegraphics[width=7.5cm]{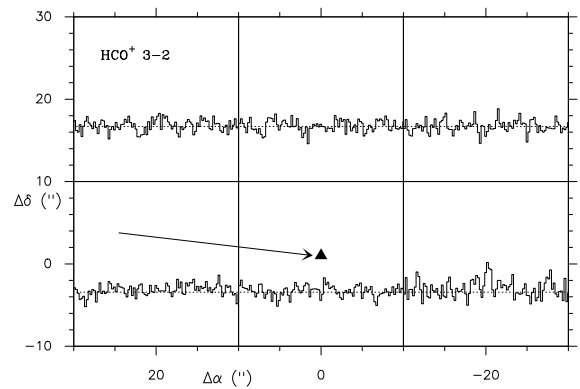}
\caption{JCMT Spectra of the HCO$^+$ (3--2) emission surrounding HH 337A. The
arrow line shows the tentative exciting source suggested by
\cite{Bally96b}.
(Note: HH 337A should not be confused with HH 337 discovered by
\citet{Noriega-Crespo02}, which  is associated with the Cep E molecular).
}
\label{337}
\end{figure}

{\it HH~337A:} This is an HH extended in the north-south about $6'$ west of 
the HH 333 jet in NGC 1333 \citep{Bally96b}. HH 337A may be part of this HH
system, in which case the exciting source would be the 333 star, but this
remains to be confirmed. Of the six positions observed, only one shows
emission, which appears to be located ahead of the HH 337, if this object
belongs to the HH 333 jet (Fig.~\ref{337}). Thus, HH~337 is a good candidate
for compact emission.

{\it HH~362A:} First presented in \citet{Alten97}.  One of a pair of faint
diffuse HH objects in the L1489 cloud, located 0.2~pc from IRAS~04106+2610.  No
detections were made.

{\it HH~366:} Driven by the protostar B5-IRS~1 in the parent dark cloud Barnard
5, HH~366 is a parsec scale bipolar outflow discovered by \citet{Bally96a}. It
is 2.2~pc in length at a distance of 350~pc.  HH~366 shows emission in HCO$^+$
3--2 in the only position observed. We observed ahead of knot E2

\begin{figure}
\centering
\includegraphics[width=3.5cm]{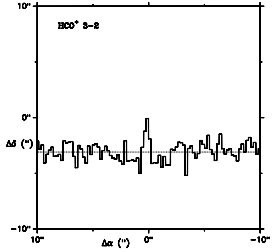}
\caption{JCMT Spectra of the HCO$^+$ (3--2) emission toward HH~366. }
\label{366}
\end{figure}

\begin{figure} 
\centering
\includegraphics[width=6cm]{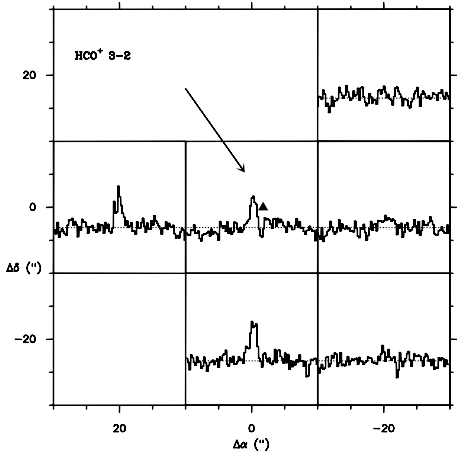}
\caption{JCMT Spectra of the HCO$^+$ (3--2) emission surrounding HH 427.
The arrow marks the location of the
nearest possible source, IRAS 03271+3013, though it unlikely to be
responsible \citep{Yan98}.}º
\label{hh427}
\end{figure}

{\it HH~427:} this object was first presented by \citet{Yan98} and it is 
located in the Barnard 1 Dark Cloud. The closest source is IRAS~03271+3013,
however its bipolar outflow extends from the NE to SW whilst HH~427 is located
0.08~pc SE of the YSO. As such, and with no other likely sources, the exciting
source remains unknown. The HCO$^+$ is detected in three positions around HH
427 (Fig.~\ref{hh427}), with the three positions west to HH 427  showing no
emission. The lower limit for the size of the HCO$^+$ emission is about
0.05~pc. Assuming that the HCO$^+$ is ahead of HH 427, then Figure~\ref{hh427}
suggests that the exciting source should be located roughly north or west of HH
427. 

{\it HH~462:} Discovered by \citet{Aspin00},  it is a small outflow along the
symmetry axis of cometary nebulae PP~11 and illuminated by the embedded
IRAS~03507+3801 source just 0.01~pc distant.   HH~462 shows emission in HCO$^+$
(3--2)  extending north and east of the HH object, spatially coincident with
the center of the optical nebulosity where the YSO is embedded. All emission
seen has a double peaked structure. All these indicate that the emission may be
tracing the dense circumstellar envelope around the known embedded Class I
source  \citep{Aspin00}. As such it has been excluded from the results
presented.

\begin{figure} 
\centering
\includegraphics[width=7.5cm]{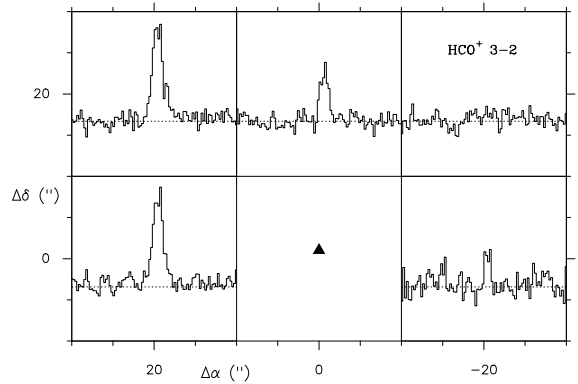}
\caption{JCMT Spectra of the HCO$^+$ (3--2) emission surrounding HH~462.}
\label{462}
\end{figure}

{\it HH~464:} Discovered by \citet{Aspin00}, HH 464 is made up of a curved
chain of 5 HH knots It appears to be powered by the T Tauri star  PP~13N, as
the chain is roughly perpendicular to the expected outflow from PP 13S, which
is responsible for the nearby HH~463. We observed downstream of knot E, which
is 0.07~pc ($40''$) south from PP~13N.   We find emission extending over a
region of 0.1 by 0.1~pc,
strong HCO$^+$ (3--2) lines are seen in every offset, this only defines the
minimum size of the region to be 0.14~pc. Interestingly, the two strongest
lines are separated by 0.07~pc, with weaker emission seen in the other
offsets. 

\begin{figure}
\centering
\includegraphics[width=7.5cm]{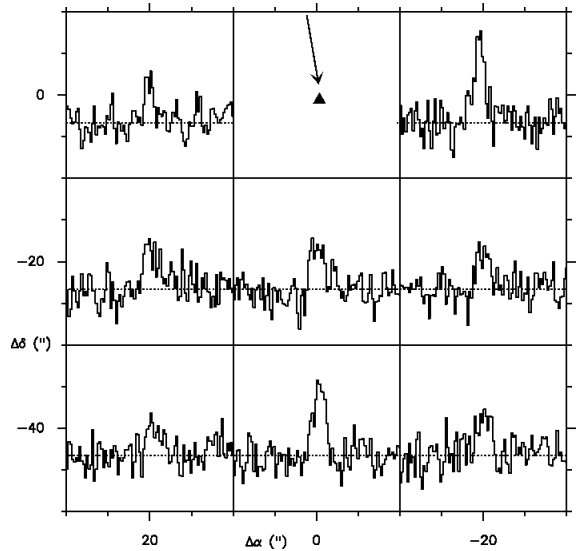}
\caption{JCMT Spectra of the HCO$^+$ 3--2 emission surrounding HH 464.}
\label{464}
\end{figure}

\begin{figure} 
\centering
\includegraphics[width=7.5cm]{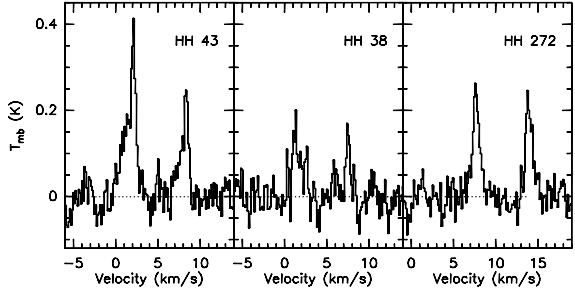}
\caption{Spectra of the CH$_3$OH (2$_n$-1$_n$) ahead of HH~43, HH~ 38 and HH~
272, observed with SEST. The offset positions for these spectra are
($0''$,$-50''$), ($0''$,$10''$) and ($0''$,$0''$), respectively.} 
\label{fig_metanol}
\end{figure}

\begin{table}
\caption{HCO$^+$ (3--2) survey results from both the JCMT and SEST }
\label{jcmt}
\centering
\begin{tabular}{l@{\hspace{0.1cm}}r@{\hspace{0.1cm}}c@{\hspace{0.1cm}}rcc}
\hline\hline
Object 
& \multicolumn{2}{c}{Positions$^a$} 
& Area$^b$ & $\!\!\!\!$T$_{mb}$ & $\Delta V$    \\ 
& \multicolumn{1}{c}{Obs} 
& \multicolumn{1}{c}{Det}     
& (pc$^{2}$)
& $\!\!\!\!$(K) &
(km s$^{-1}$) \\
 \hline
HH~29	&  3 & 3 &   $4.3\times10^{-4}$ 		& 0.6$\pm$0.1 & 1.8$\pm$0.2 \\
HH~38  	& 11 & 1 &  $1.7\times10^{-2}$ 	& 1.2$\pm$0.3 & 0.9$\pm$0.2  \\ 
HH~43	&  4 & 1 &  $6.3\times10^{-3}$ 		& 0.6$\pm$0.2 & 0.7$\pm$0.2  \\
HH~47C	& 11 & 0 & $1.7\times10^{-2}$		& $<1.4$      & \nodata   \\
HH~49	& 22 & 0 & $4.2\times10^{-3}$ 		& $<0.6$      & \nodata   \\  
HH~52-53&  9 & 0 &  $1.8\times10^{-3}$		& $<2.4$      & \nodata   \\
HH~54	& 12 & 0 &  $2.4\times10^{-3}$	& $<0.7$      & \nodata   \\
HH~75	&  9 & 0 &  $1.4\times10^{-2}$		& $<0.8$      & \nodata   \\
HH~77   &  9 & 4 & 	$3.3\times10^{-2}$	& 1.1$\pm$0.3 & 1.6$\pm$0.2  \\
HH~211  &  6 & 6 &  $4.0\times10^{-3}$		& 1.2$\pm$0.1 & 1.4$\pm$0.1  \\
HH~240  & 11 & 3 &  $1.7\times10^{-2}$		& 1.4$\pm$0.3 & 0.6$\pm$0.1  \\
HH~267 	&  6 & 0 & $4.0\times10^{-3}$ 		& $<0.38$     & \nodata   \\
HH~268 	&  3 & 0 &  $2.0\times10^{-3}$ 		& $<0.56$     & \nodata   \\
HH~272  &  9 & 3 & 	$4.6\times10^{-2}$	& 1.1$\pm$0.4 & 0.8$\pm$0.3  \\
HH~276 	&  6 & 0 &  $8.7\times10^{-4}$ 		& $<0.53$     & \nodata   \\
HH~278	&  6 & 0 &   $4.0\times10^{-3}$		& $<0.43$     & \nodata   \\
HH~337A &  6 & 1 &   $4.0\times10^{-3}$		& 0.5$\pm$0.2 & 0.5$\pm$0.1  \\
HH~362A &  6 & 0 &   $8.7\times10^{-4}$		& $<0.46$     & \nodata   \\
HH~366 	&  1 & 1 &   $ 6.7\times10^{-4}$		& 0.5$\pm$0.1 & 0.3$\pm$0.1  \\
HH~427 	&  6 & 3 &  $4.0\times10^{-3}$		& 0.9$\pm$0.1 & 0.9$\pm$0.1  \\
HH~462 	&  5 & 4 &   $3.3\times10^{-3}$		& 2.0$\pm$0.2 & 1.0$\pm$0.1  \\
HH~464 	&  8 & 8 &   $7.2\times10^{-3}$		& 1.1$\pm$0.1 & 1.2$\pm$0.1  \\
\hline
\end{tabular}
\begin{list}{}{}
     \item[$^{a}$] 
Obs: n$^{\rm o}$ of positions observed, Det: n$^{\rm o}$ of positions
with $\geq 3\sigma$ detections.     
     \item[$^{b}$] 
The area covered by the observations around the HH object 
\end{list}
\end{table}

\begin{table}
\caption{H$^{13}$CO$^+$ (1--0) survey results from SEST.}
\label{sest}
\centering
\begin{tabular}{l r r c c}
\hline\hline
Object 
& \multicolumn{2}{c}{Positions$^a$} 
& T$_{mb}$& $\Delta V$  \\
& \multicolumn{1}{c}{Obs} 
& \multicolumn{1}{c}{Det}     
&  (K)    & (km s$^{-1}$)\\
 \hline
HH~38  		&   11 	& 3 	& 0.28$\pm$0.05	& 0.9$\pm$0.1 \\
HH~43		&    4 	& 1 	& 0.21$\pm$0.04 	& 0.6$\pm$0.1 \\
HH~47C		&   11 	& 0 	& $<0.23$ 		& \nodata \\
HH~49		&   22 	& 0 	& $<0.19$ 		& \nodata \\
HH~52-53		&    9 	& 0 	& $<0.26$ 		& \nodata \\
HH~54		&   12 	& 0 	& $<0.22$ 		& \nodata \\
HH~75		&    9 	& 0 	& $<0.17$ 		& \nodata \\ 
HH~77$^b$	&   9 	& 4 	& 0.34$\pm$0.04 	& 1.2$\pm$0.1 \\  
HH~240  		&   11 	&10 	& 0.45$\pm$0.06 	& 0.6$\pm$0.1 \\
HH~272  		&    9 	& 5 	& 0.33$\pm$0.06 	& 0.8$\pm$0.1 \\
\hline
\end{tabular}
\begin{list}{}{}
     \item[$^{a}$] 
Obs: n$^{\rm o}$ of positions observed, Det: n$^{\rm o}$ of positions
with $\geq 3\sigma$ detections.     
     
     \item[$^{b}$]
HH~77 has an additional weak, 2~$\sigma$ detections, in other two positions.
\end{list}
\end{table}

\subsection{Survey results}

Table~\ref{jcmt} details the HCO$^+$ (3--2) results for all sources. Column 2
lists for each source, the total number of offsets in which emission was
observed in the region surrounding the HH object. Column 3 lists the number 
of offsets in which emission was seen. Column 4 indicates the total area we 
mapped for emission. A full mapping of the suspected region for each
source was not possible due to time constraints and the differing area on the
sky covered by the 20$''$ beam for sources at different distances. 
In addition as the orientation of the exciting source and the direction of 
both the outflow and the UV illumination are not known with certainty, and as 
such the region being observed is not always located exactly ahead of the  HH object.
The remaining columns list the peak main beam temperature (5) and the 
half-power line width (6). The SEST observation of H$^{13}$CO$^+$ (1--0) 
is similarly detailed in Table~\ref{sest}. For non-detections, upper limits on the 
line intensities are given by 
$I(\rm {HCO^{+}}) < 3\sigma_{\rm mb}$ where $\sigma_{\rm mb}$ is the main 
beam RMS noise of the spectrum.

Eleven out of the twenty one objects surveyed (excluding HH~462 which seems to
show emission from the dense circumstellar envelope around the exciting YSO),
52$\%$, had HCO$^+$ (3--2) emission observed in at least one offset from
the Herbig-Haro object.  The HH objects observed were situated between 
0.07 and 2.7~pc (projected distances) from their powering sources. 
The HCO$^+$ (3--2) emission associated with
the dense circumstellar envelope around low mass protostars arises from a
region of $\sim0.02$~pc in diameter \citep{Hogerheijde97}. Thus, we do not
expect to be contaminated from the circumstellar molecular component. Of the
sample no source further than 1~pc away from its powering source produced
emission in the surrounding areas. The quiescent region we observed was
typically between 0.01--0.07~pc from the HH object. The size of the
emitting regions has been estimated from the number of offset positions that
see emission (with a beam size of 20$''$ for HCO$^+$ (3--2)), although it is
extremely important to note that emission may extend beyond the regions mapped,
as in a number of cases emission is seen up to the edge of the observed region.
Five of the regions with several positions observed show detection in only few
positions (HH 38, HH 43, HH 240 and HH 337). This suggests that the
emission comes from compact clumps, probably with diameter between 
$\sim0.03$ to 0.07~pc. In HH 77 and HH 427 there is some evidence that the 
emission may be also compact (about half of the positions observed had 
HCO$^+$ emission). HH 211 and HH 464 show emission in all the positions 
observed, implying an emission scale of $\sim0.1$~pc or larger. 
For the SEST observations, if HCO$^+$ (3--2) was seen, then in all cases
H$^{13}$CO$^+$ (1--0) was also seen.  In HH 240 and HH 38 the H$^{13}$CO$^+$
(1--0) emission is more extended than the HCO$^+$ (3-2) line. Because of the
lower critical density for the lower J rotational transition, this is an
indication that the H$^{13}$CO$^+$ line is tracing lower density gas.

\section{Discussion and conclusions}

\subsection{The origin of the emission}

\begin{figure} 
\centering
\includegraphics[width=7.5cm]{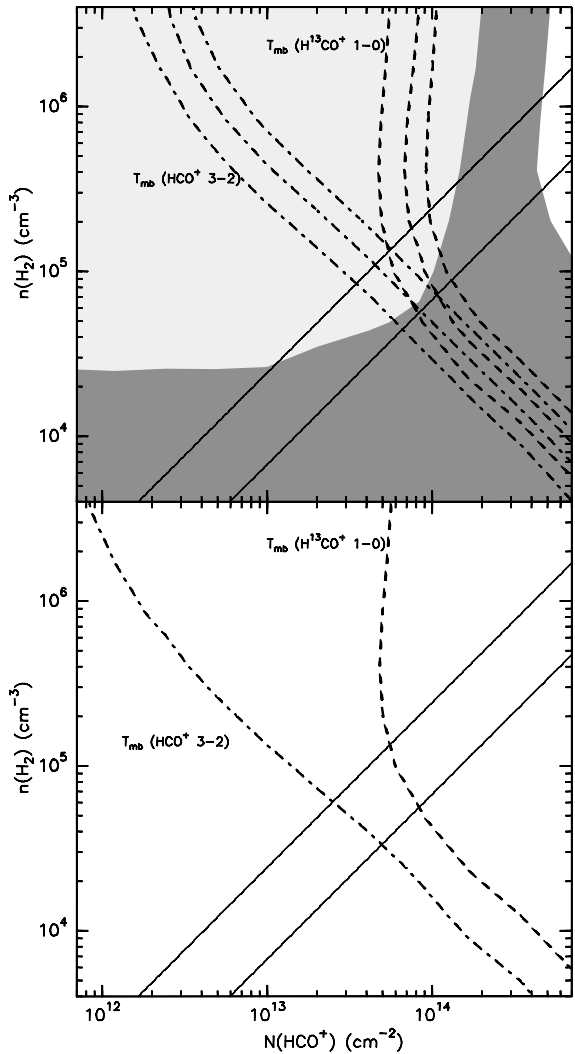}
\caption{
{\em Top panel:} Plot of the set of RADEX solutions in the 
$N$(HCO$^+$)--$n$(H$_2$ ) plane for  HH 38 assuming a kinetic temperature 
of 10~K and a Gaussian source with a FWHM of $30''$ or 0.07~pc.  The 
dotted-dashed lines show the range of solutions for the observed intensity of the 
HCO$^+$ 3--2 line (including the uncertainties and taking into account the filling 
factor for the assumed size).  The dashed lines shows the solutions for the 
H$^{13}$CO$^+$ 1--0 line. The grey area is the range of solutions for the 
observed line ratio for these two lines. The solid line shows the range of solutions 
for equation~\ref{eqn}, for an HCO$^+$ abundance in the 2 to $5 \times 10^{-9}$ 
range.
{\em Bottom panel:} Similar to the top panel but for an observed intensity of 0.5 
and 0.2~K for the HCO$^+$ 3--2 and H$^{13}$CO$^+$ 1--0 line, respectively. 
Higher values of the line intensities should appear above these lines.}
\label{radex}
\end{figure}

The targets with detected HCO$^+$ (3--2) emission have sizes of about $\la
0.1$~pc, except for those where the coverage was not enough to constrain the
size (see previous section).  However, this alone is not enough to state that
the emission is coming from dense and small clumps or condensations within the
dark molecular cloud.  In order to investigate if this is the case, we derived
the density of the  molecular gas traced by the HCO$^+$  
for HH~38, a region where we have the two isotopologues and that shows
compact emission.  
The gas volume density can be estimated in the following way:
\begin{equation}\label{eqn}
n(\mathrm{H_2})\simeq\frac{N(\mathrm{HCO^+})}{X(\mathrm{HCO^+})}\,\frac{1}{L},
\end{equation}
where $L$ is the scale length of the emission, and $N(\mathrm{HCO^+})$ and 
$X(\mathrm{HCO^+})$ are the average values for the  clump of the HCO$^+$ column
density and the HCO$^+$  abundance relative to H$_2$, respectively.  
We adopt a length scale of  $L=0.07$~pc, which is roughly the diameter found 
for the HCO$^+$ (3--2) emission in HH~38. We adopt an HCO$^+$ abundance 
from  2 to $5 \times 10^{-9}$, values similar to those found in other HH
objects \citep{Girart05,Viti06}.  
We used the RADEX code \citep{Tak07}, a non-LTE molecular radiative transfer in
an isothermal homogeneous medium, to derive the HCO$^+$ (3--2) 
and H$^{13}$CO$^+$ (1--0) line intensities for a kinetic temperature of
10~K and for the adopted length scale (to include the filling factor correction
in the derived RADEX intensities).
As shown in Figure~\ref{radex}, the observed line intensities of HH~38 can be 
accounted for densities roughly  near $10^5$~cm$^{-3}$.
In addition, if we consider that the detection threshold of our observations
are $\simeq0.5$~K and 0.2 for the HCO$^+$ (3--2) and H$^{13}$CO$^+$ (1--0),
respectively, then for the same aforementioned conditions, the density of the
detected emission should be higher than
$n(\mathrm{H_2})\ga3\times10^4$~cm$^{-3}$ (see Fig.\ref{radex}).

In any case, the lower limit derived for the volume density is clearly higher
than the average density in molecular dark clouds, expected to be 
$\sim10^3$~cm$^{-3}$  (e.g. \citet{Loehr07}).  In addition, it should be noted
that for most of the  detections, the emission is quite far from the powering
source of the HH object,   so the contribution of the dense core around the YSO
to the derived volume density is negligible.  We conclude therefore that the
bulk of the observed emission is indeed arising from small, compact, dense
clumps with a minimum density of 
$3\times10^4$~cm$^{-3}$.

\subsection{Estimation of clumpiness}
It may be possible, at least in principle, to make an estimate of the 
clumpiness of molecular clouds from observations of HCO$^{+}$ emission in the 
vicinity of Herbig-Haro objects. Here, we indicate one method by which this 
estimate might be made. Of course, there are several caveats that 
must be accepted in attempting to do this. Firstly, the observed HCO$^{+}$ 
emission may be misleading in the sense that molecular clouds may contain 
filamentary or sheet-like structures that only appear to be clumpy because 
of their morphology; in such a case, the clumps may have no real 
significance. Secondly, the high-density emitting gas described in this 
paper may not be representative of the entire cloud, but only of a small 
region close to the HH object.

At present, there is rather limited evidence from high spatial resolution 
studies of the structure of molecular clouds. \citet{Morata03, Morata05} 
made interferometric molecular line observations of a restricted region (about 
0.7 pc $\times$ 0.7 pc) of the molecular cloud L673. When combined with single-dish 
observations of the same region, the observations indicated that this 
limited region is populated by 15 resolved and distinct clumps of 
diameters less than about 0.1 pc and with gas number densities larger than 
$1\times10^4$~cm$^{-3}$. The clumps are transient and show evidence of 
time-dependent chemistry. Peng et al. (1998) and \citet{Takakuwa03} found 
similar results for a region of the molecular cloud TMC-1. 

These observations are limited to small regions of molecular clouds. As discussed 
in Section 1, there is, however, indirect evidence that supports the idea that 
clumps may be widespread in molecular clouds \citep{Garrod06}.

Even if subsequent observations show that clumps are confined to restricted 
regions of molecular clouds, the distribution of matter in those regions is of 
interest, since the denser parts of the gas have the potential for low-mass star 
formation. We now proceed to make estimates of clumpiness on the 
assumption that discrete transient clumps are present in the vicinity of 
an HH object that is accompanied by HCO$^+$ emission. Whether or not the inferred 
measures of clumpiness should be applied to the entire cloud, in each 
case, will depend on the outcome of future interferometric observations of 
molecular clouds.

From our survey, one can derive a very rough statistical measure of clumpiness by
considering the number of HCO$^+$ detections: the first consideration to make
is that of the 21 objects observed (excluding HH 462) we had a positive
detection in at least one position for $\sim$ 50\% of the objects implying that, 
roughly, one in two HH objects is moving through a clumpy medium. Note that 
our survey probably did not cover all the regions that may emit strong HCO$^+$ 
for each object, hence this estimate may in fact be a lower limit. 

\citet{Taylor96} suggested that an estimate of the mean linear separation
between clumps detected near HH objects is given by
\begin{equation} 
\left[\frac{l}{\mathrm{cm}}\right] = 10^{17} 
\left[\frac{v}{30 \, \mathrm{km \,s^{-1}}}\right]
\left[\frac{t}{100 \, \mathrm{yrs}}\right]
\left[\frac{f}{0.1}\right]^{-1}
\end{equation}
where $v$ is the velocity of the HH object, $t$ is the duration of the anomalous
emission from the irradiated clump, and $f$ is the fraction of jets that have
these clumps associated with them. For present purposes we adopt equation (2). 
The value of $t$ is unconstrained in equation (2). We make a crude estimate of $t$ by arguing that it is the time for the HH object to travel a distance equivalent to the clump diameter. For the clump size of $\sim$ 0.07 pc as measured for HH 38 (see Section 4.1) and for the canonical velocity in equation (2) this gives a value of $t$ of roughly 2$\times$10$^3$ yrs, that is consistent with a detailed dynamical study of the time evolution of clump chemistry (Christie et al. submitted).
With $t=2r/v$, then equation (2) gives $l=2r/f$. For the clump diameter of HH 38, $l$ is then 0.14pc. An estimate of the area filling factor, $f_A$, is 
$(r/l)^2$, i.e. $(f/2)^2$. with $f$ = 0.5, then $f_A$ = 6\% for this example. Of course, this estimate is uncertain, but it is clear that an estimate can be made along these lines. An estimate of $f_A$ from the observational data of Morata et al. (2005) suggests that $f_A$ in L673 should be about 30\%.

\subsection{Conclusions}

We have performed a large scale survey of HCO$^+$ (3--2)  ahead of a sample of
22 sources, where no previous HCO$^+$ emission had been seen. Ten sources
showed emission in HCO$^+$ (3--2) and when H$^{13}$CO$^+$ (1--0) observations
were possible emission was also always seen.  These clumps have velocities in
line with the V$_{LSR}$ of the cloud, narrow linewidths of $\la~1$~km s$^{-1}$
and hence they seem dynamically unaffected by the jet. The gas number densities
of these clumps are close to 2 orders of magnitude higher than the mean density
of molecular clouds ($n({\rm H_2}) \sim 10^5$~cm$^{-3}$ for the case of the 
HH 38 region); this implies that these clouds may be highly heterogeneous
at scales of less than 0.1~pc. 

Our attempt at studying clumpiness by the use of HCO$^+$ demonstrates that  it
is possible, in principle, to use Herbig-Haro objects as probes of clumpiness in molecular
clouds.  Ultimately, future combined mm (mainly 3~mm)  interferometric and
single dish observations at relatively large scales with high sensitivity are
required to fully determine the mass spectrum of the clumps, in a similar way
of the work done by Morata et al. (2005).  While we do not here attempt any
explanation for the origin of molecular clumpiness, if slow-mode MHD waves
\citep{Falle02} and MHD turbulence \citep{Ballesteros07} are responsible for
this highly heterogeneous medium, this mechanism should be such that within the
clumps the non thermal linewidth is of the order or less than 1~km~s$^{-1}$.

Our conclusions regarding measures of clumpiness should be qualified 
by the following remarks. The detection of enhanced HCO$^+$ emission 
close to the HH object enables us to estimate clumpiness only 
in the regions where high density gas is present, and may not 
represent the clumpiness throughout the cloud. 
Other observational approaches (e.g. Morata et al. 2003, 2005; Peng et al. 1998) detect clumpiness, but their studies are also of relatively 
small regions of molecular clouds. We cannot, therefore, 
conclude that the clumpiness estimate we have made is applicable
to the entire molecular cloud. However, we note a modelling approach 
(Garrod et al. 2006) suggests that the clumpiness should be found throughout the molecular cloud.

Finally, it is implicit throughout this work that the clumpy 
structure is of essentially discrete parcels of relatively dense gas. 
This may not be the case; it is possible that the 
denser gas is in filaments rather than discrete clumps. 
The highest resolution maps of molecular clouds available 
(Morata et al. 2005 for L673) do seem to suggest separate structures, 
but of course unresolved material may also be present. 
Our assumption of discrete clumps is, therefore, plausible, 
but our estimates would have to revised if further evidence 
supports the filamentary nature of cloud structure.
\begin{acknowledgements} 
The James Clerk Maxwell Telescope is operated by The Joint Astronomy Centre on
behalf of the Particle Physics and Astronomy Research Council of the United
Kingdom, the Netherlands Organisation for Scientific Research, and the National
Research Council of Canada. The SEST is operated jointly by the European
Southern Observatory and the Swedish National Facility for Radio Astronomy,
Onsala Space Observatory at Chalmers University of Technology. WW is supported
by a PPARC studentship. JMG and RE are supported by the Ministerio de Ciencia e 
Innovaci\'on AYA2008-06189-C03 grant. 
JMG, SV and RE acknowledge support by a joined Royal Society and CSIC travel grant.
\end{acknowledgements}


\begin{thebibliography}{}
\bibitem[\protect\citeauthoryear{Alten et al.}{1997}]{Alten97} 
 Alten V.~P., Bally J., Devine D., Miller G.~J., 1997, IAUS, 182, 51P 
\bibitem[\protect\citeauthoryear{Aspin \&  Reipurth}{2000}]{Aspin00}
 Aspin C., Reipurth B., 2000, \mnras, 311, 522 
\bibitem[Ballesteros-Paredes et al.(2007)]{Ballesteros07} 
 Ballesteros-Paredes, J., Klessen, R.~S., Mac Low, M.-M., \& Vazquez-Semadeni,
 E.\ 2007, Protostars and Planets V, 63 
\bibitem[\protect\citeauthoryear{Bally, Devine, \& Alten}{1996}]{Bally96a}
 Bally J., Devine D., Alten V., 1996, \apj, 473, 921 
\bibitem[\protect\citeauthoryear{Bally, Devine, \& Reipurth}{1996}]{Bally96b}
 Bally J., Devine D., Reipurth B., 1996, \apj, 473, L49 
\bibitem[\protect\citeauthoryear{Bally et al.}{1997}]{Bally97} 
 Bally J., Devine D., Alten V., Sutherland R.~S., 1997, \apj, 478, 603 
\bibitem[\protect\citeauthoryear{Bally et al.}{2006}]{Bally06} 
 Bally J., Walawender J., Luhman K.~L., Fazio G., 2006, AJ, 132, 1923 
\bibitem[\protect\citeauthoryear{Carballo \&  Eiroa}{1992}]{Carballo92}
 Carballo R., Eiroa C., 1992, \aap, 262, 295 
\bibitem[\protect\citeauthoryear{Cohen}{1980}]{Cohen80}
 Cohen M., 1980, AJ, 85, 29 
\bibitem[\protect\citeauthoryear{Cohen}{1990}]{Cohen90} 
 Cohen M., 1990, \apj, 354, 701 
\bibitem[\protect\citeauthoryear{Davis et al.}{1997}]{Davis97} 
 Davis C.~J., Ray T.~P., Eisloeffel J., Corcoran D., 1997, \aap, 324, 263 
\bibitem[Dent et al.(1993)]{Dent93}
 Dent, W.~R.~F., Cunningham, C., Hayward, R., et al.\ 1993, \mnras, 262, L13
\bibitem[\protect\citeauthoryear{Eisl{\"o}ffel \& Mundt}{1998}]{Eisloffel98}
 Eisl{\"o}ffel J., Mundt R., 1998, AJ, 115, 1554 
\bibitem[\protect\citeauthoryear{Eisl{\"o}ffel}{2000}]{Eisloffel00} 
 Eisl{\"o}ffel J., 2000, \aap, 354, 236 
\bibitem[\protect\citeauthoryear{Falle \& Hartquist}{2002}]{Falle02}
 Falle, S.~A.~E.~G., \& Hartquist, T.~W.\ 2002, \mnras, 329, 195 
\bibitem[\protect\citeauthoryear{Garrod et al.}{2006}]{Garrod06}
 Garrod, R.~T., Williams, D.~A., \& Rawlings, J.~M.~C.\ 2006, \apj, 638, 827 
\bibitem[\protect\citeauthoryear{Girart et al.}{2000}]{Girart00}
 Girart, J.~M., Estalella, R., Ho, P.~T.~P., \& Rudolph, A.~L.\ 2000, \apj,
 539, 763 
\bibitem[\protect\citeauthoryear{Girart et al.}{2005}]{Girart05} 
 Girart J.~M., Viti S., Estalella R., Williams D.~A., 2005, \aap, 439, 601 
\bibitem[\protect\citeauthoryear{Girart et al.}{2002}]{Girart02} 
 Girart J.~M., Viti S., Williams D.~A., Estalella R., Ho P.~T.~P., 2002, 
 \aap, 388, 1004 
\bibitem[\protect\citeauthoryear{Haro}{1953}]{Haro53}
 Haro G. 1953, \apj, 117, 73
\bibitem[\protect\citeauthoryear{Herbig}{1974}]{Herbig74}
 Herbig G.~H., 1974, LicOB, 658, 1 
\bibitem[\protect\citeauthoryear{Herbst}{1975}]{Herbst75}
 Herbst W., 1975, AJ, 80, 212 
\bibitem[\protect\citeauthoryear{Hodapp \& Ladd}{1995}]{Hodapp95}
 Hodapp K.-W., Ladd E.~F., 1995, \apj, 453, 715 
\bibitem[\protect\citeauthoryear{Hogerheijde et al.}{1997}]{Hogerheijde97}
 Hogerheijde, M.~R., van Dishoeck, E.~F., Blake, G.~A., \& van Langevelde,
 H.~J.\ 1997, \apj, 489, 293 
\bibitem[\protect\citeauthoryear{Knee}{1992}]{Knee92}
 Knee, L. B. G. 1992, \aap, 259, 283
\bibitem[\protect\citeauthoryear{Langer \& Penzias}{1993}]{Langer93}
 Langer, W.~D., \& Penzias, A.~A.\ 1993, \apj, 408, 539
\bibitem[\protect\citeauthoryear{Lefloch et al.}{2005}]{Lefloch05}
 Lefloch, B., Cernicharo, J., Cabrit, S., \& Cesarsky, D.\ 2005, \aap, 433,
 217 
\bibitem[\protect\citeauthoryear{Loehr et al.}{2007}]{Loehr07}
 L{\"o}hr, A., Bourke, T.~L., Lane, A.~P., Myers, P.~C., Parshley, S.~C., 
 Stark, A.~A., \& Tothill, N.~F.~H.\ 2007, \apjs, 171, 478
\bibitem[\protect\citeauthoryear{McCaughrean, Rayner, \& Zinnecker}{1994}]{McCaughrean94}
 McCaughrean M.~J., Rayner J.~T., Zinnecker H., 1994, \apj, 436, L189 
\bibitem[\protect\citeauthoryear{Morata, Girart, \& Estalella}{2003}]{Morata03}
 Morata O., Girart J.~M., Estalella R., 2003, \aap, 397, 181 
\bibitem[\protect\citeauthoryear{Morata, Girart, \& Estalella}{2005}]{Morata05}
 Morata O., Girart J.~M., Estalella  R., 2005, \aap, 435, 113 
\bibitem[\protect\citeauthoryear{Moriarty-Schieven et al.}{2006}]{Moriarty06}
 Moriarty-Schieven G.~H., Johnstone D., Bally J., Jenness T., 2006, \apj, 645,
 357 
\bibitem[\protect\citeauthoryear{Noriega-Crespo}{2002}]{Noriega-Crespo02}
 Noriega-Crespo A., 2002, RMxAC, 13, 71 
\bibitem[\protect\citeauthoryear{Peng et al.}{1998}]{Peng98}
 Peng, R., Langer, W.~D., Velusamy, T., Kuiper, T.~B.~H., \& Levin, S.\ 1998,
 \apj, 497, 842 
\bibitem[\protect\citeauthoryear{Raga \& Williams}{2000}]{Raga00}
 Raga, A.~C., \& Williams, D.~A.\ 2000, \aap, 358, 701 
\bibitem[\protect\citeauthoryear{Reipurth \& Graham}{1988}]{Reipurth88}
 Reipurth B., Graham J.~A., 1988, \aap, 202, 219 
\bibitem[\protect\citeauthoryear{Reipurth, Nyman, \& Chini}{1996}]{Reipurth96}
 Reipurth B., Nyman L.-A., Chini R., 1996, \aap, 314, 258 
\bibitem[\protect\citeauthoryear{Rudolph \& Welch}{1988}]{Rudolph88}
 Rudolph, A., \& Welch, W.~J.\ 1988, \apjl, 326, L31 
\bibitem[\protect\citeauthoryear{Rudolph \& Welch}{1992}]{Rudolph92}
 Rudolph, A., \& Welch, W.~J.\ 1992, \apj, 395, 488 
\bibitem[\protect\citeauthoryear{Sahu, Sahu \& Pottasch}{1989}]{Sahu89} 
 Sahu, M., Sahu, K. C., \& Pottasch, S. R.\ 1989, \aap, 218, 221
\bibitem[\protect\citeauthoryear{Schwartz}{1977}]{Schwartz77b} 
 Schwartz R.~D., 1977, \apj, 212, L25 
\bibitem[\protect\citeauthoryear{Schwartz}{1977}]{Schwartz77a} 
 Schwartz R.~D., 1977, ApJS, 35, 161 
\bibitem[\protect\citeauthoryear{Schwartz}{1978}]{Schwartz78} 
 Schwartz R.~D., 1978, \apj, 223, 884 
\bibitem[\protect\citeauthoryear{Stanke et al.}{2000}]{Stanke00}
 Stanke, T., McCaughrean, M.~J., \& Zinnecker, H.\ 2000, \aap, 355, 639 
\bibitem[\protect\citeauthoryear{Takakuwa et al.}{2003}]{Takakuwa03}
 Takakuwa S., Kamazaki T.,  Saito M., Hirano N., \apj, 584, 818
\bibitem[\protect\citeauthoryear{Taylor \& Williams}{1996}]{Taylor96}
 Taylor, S. D. \& Williams D.~A., 1996, \mnras, 282, 1343
\bibitem[\protect\citeauthoryear{Torrelles et al.}{1992}]{Torrelles92}
 Torrelles, J.~M., Rodriguez, L.~F., Canto, J., et al.\ 1992, \apjl, 396, L95 
\bibitem[\protect\citeauthoryear{van der Tak et al.}{2007}]{Tak07}
 van der Tak, F.~F.~S., Black, J.~H., Sch{\"o}ier, F.~L., Jansen, D.~J., \& van
 Dishoeck, E.~F.\ 2007, \aap, 468, 627 
\bibitem[\protect\citeauthoryear{Viti et al.}{2003}]{Viti03} 
 Viti S., Girart J.~M., Garrod R., Williams D.~A., Estalella R., 2003, \aap,  
 399, 187 
\bibitem[\protect\citeauthoryear{Viti, Girart, \& Hatchell}{2006}]{Viti06}
 Viti S., Girart J.~M., Hatchell J., 2006, \aap, 449, 1089 
\bibitem[\protect\citeauthoryear{Wu et al.}{2002}]{Wu02}
 Wu J.-W., Wu Y.-F., Wang J.-Z., Cai K., 2002, ChJAA, 2, 33 
\bibitem[\protect\citeauthoryear{Yan et al.}{1998}]{Yan98}
 Yan J., Wang H., Wang M., et al. 1998, AJ, 116, 2438 
      
\end{thebibliography}
\end{document}